# Detecting Musical Deepfakes

Nicholas Sunday

*Department of Computer Science, The University of Texas at Austin, USA*

nsunday@utexas.edu

*Abstract*—The proliferation of Text-to-Music (TTM) platforms has democratized music creation, letting users effortlessly generate high-quality compositions. However, this innovation has also introduced challenges to musicians and the music industry. This research focuses on utilizing the FakeMusicCaps dataset to address the challenge of detecting AI-generated songs by classifying the audio as deepfake or human. To simulate a real-world adversarial entity tempo stretching and pitch shifting modifications were applied to the dataset. Mel Spectrograms were generated from the resulting datasets, which were then used to train and test a convolutional neural network. This paper also explores the ethical and societal implications of TTM platforms, suggesting that detection systems developed and employed with care are a necessary tool to safeguard musicians and foster the positive potential of TTM platforms and generative AI in music.

## I. INTRODUCTION

Rapid advances in generative AI have caused the creative landscape to be upended, enabling almost anyone to easily create music that can be hard to distinguish from human-made compositions. AI-generated music is part of a wider classification of AI-generated media and art that falls under the category of "deepfake". The term "deepfake" is a portmanteau of the terms "deep learning", from the AI models used to generate the art, and "fake", refers to the non-human-made nature. In recent years many Text-to-Music (TTM) platforms have made this technology even more accessible. While these platforms, and the music they create, open new avenues for artistic exploration and accessibility, they also present legal, ethical, and technical challenges. Copyright infringement, fraudulent attribution, and the erosion of artistic authenticity are some examples of the issues presented by the proliferation of TTM deepfake platforms. These issues necessitate deepfake detection systems that are capable of distinguishing between human music and AI-generated tracks.

In this research, I aim to contribute research to the growing field of deepfake detection by utilizing the FakeMusicCaps dataset to test the effectiveness of deep learning models in identifying musical deepfakes. By focusing the scope of the classification down to a binary problem, human or deepfake, I can evaluate the performance of the ResNet18 model under various conditions mimicking those a malicious actor might employ to evade detection systems.

This paper builds upon the deepfake detection work done by previous teams of researchers. The FakeMusicCaps dataset was created by researchers at Politecnio di Milano to lay a baseline foundation for future detection work. I combined this dataset with adversarial manipulation methods employed by the research team at Deezer by adding pitch shift and tempo stretch effects to the dataset. Finally, I drew inspiration from the Synthetic-Or-Not – Identifying Counterfeit Songs research team and converted my dataset to Mel Spectrograms before treating the dataset as a computer vision problem.

## II. RELATED WORKS

### A. FakeMusicCaps

A team of researchers at Italy's Politecnico di Milano set out to create a robust dataset that could be used to train future models to detect musical deepfakes as well as which popular TTM platform they were generated with [1]. Their work used the MusicCaps dataset, which is comprised of "music-text pairs, with rich text descriptions provided by human experts [2]". FakeMusicCaps builds upon this dataset by creating deepfake audio using the captions found in MusicCaps using several popular TTM platforms. After creating the dataset, their research focused on showing how FakeMusicCaps could be used with 3 popular models. For their research RawNet2 was trained using the raw audio files, and ResNet18+Spec and ResNet18 were trained on log-spectrograms.

### B. SONICS

The research team behind the Synthetic-Or-Not – Identifying Counterfeit Songs (SONICS) project, took a different approach to creating a novel dataset for advanced deepfake detection. Their dataset project aimed to create full-length artificial songs, because in their estimation previous models had limited music-lyrics diversity, short-duration songs, and lacked end-to-end fake songs. Additionally, they devised a new type of architecture, called SpecTTTra, that improves time and memory efficiency over conventional Convolutional and Transformer based models by creating clips from distinct slices of the input spectrograms. These clips are then tokenized and transformed resulting in globally contextualized features that can be averaged and then classified [3].

### C. Deezer

The team of researchers at Deezer took an interesting approach. First, they ran experiments on classifying human and deepfake audio using a convolutional model. Finding this task reasonably easy for their model, they then applied some rudimentary manipulation techniques to their songs and tested the models again. They found that their models had a harder time detecting deepfakes when manipulation had been applied, such as pitch shifting or temporal stretching [4].

## III. METHODOLOGY

### A. Dataset

The FakeMusicCaps dataset was chosen for this project because it contains real and deepfake audio. It is comprised of 5,373 human generated clips that are 10 seconds in length. Additionally, there are 5,521 10-second deepfake tracks from each of the 5 following TTM platforms: MusicGen_medium, audioldm2, musicldm, mustango, and stable_audio_open. There are also 63 longer deepfake tracks from the Suno platform.

For experimentation, all the audio clips were converted to Mel Spectrograms using the librosa python library. To select an equal number of deepfake tracks, 5,373 tracks were selected from the 5 different deepfake platform subsets of the FakeMusicCaps dataset. That gives my dataset 10,746 total tracks. For these experiments the ability of the model to detect the binary classification of the audio was the goal, so the tracks were re-classified as human or deepfake for training and testing.

### B. Model

I used ResNet18 pretrained with the default, Imagenet1k_v1, weights for these experiments, since this model has been proven to be successful at classifying images and especially Mel Spectrograms [6]. ResNet18 is an 18 layer deep convolutional neural network that is available as part of the torchvision python library. The final fully connected layer was modified to have two output features, based on the two input classes: deepfake or human. The model was trained with the Adam optimizer and cross entropy as the loss function and the for all datasets.

### C. Malicious Manipulation

For this experiment, I wanted to see how susceptible this model and method of detection was to a malicious actor. I performed two types of manipulations, pitch shifting and tempo stretching/compression. Librosa is a python library that includes methods for manipulating speed and pitch of audio files [7], and that is the method that was used to manipulate the audio files in the dataset.

For the pitch shifted dataset, I first selected a random number between -2 and +2. I then shifted the pitch of the audio clip n-semitones, where n is the random number between -2 and +2. A new random number was selected for each audio clip.

For the temporal stretched dataset, a similar random number generation was performed, this time between 0.8 and 1.2. The numbers correspond to the stretch percentage of the new clip. Using this random number, I applied a time stretch to the clip by the amount randomly generated. A different random number was generated for each audio clip.

### D. Mel Spectrograms

Four separate datasets were created for my experiments. They were then converted to Mel Spectrograms to use with ResNet18. Spectrograms are a visual representation of audio, where the horizontal axis is time, the vertical axis represents frequency, and the color intensity represents the amplitude of a frequency at a certain point in time [8]. Mel Spectrograms are Spectrograms which have been converted into the Mel Scale. The Mel Scale "is a scale of pitches, such that each unit is judged by listeners to be equal in pitch distance from the next [9]."

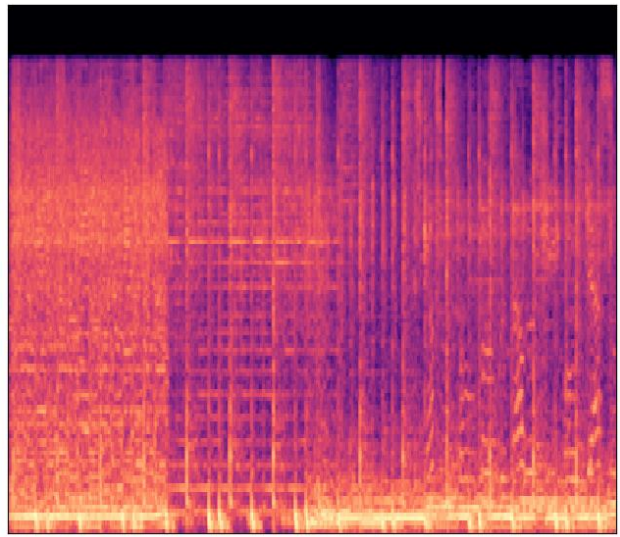

Fig. 1. A Mel Spectrogram created from a 10 second music clip found in the FakeMusicCaps dataset

## IV. EXPERIMENTS

### A. Datasets

Five sets of experiments were performed on a combination of four datasets: the base FakeMusicCaps dataset, a pitch shifted version, a tempo stretched dataset, and finally a dataset that was both pitch shifted and tempo stretched. Each dataset was modified, if necessary, and then transformed into a set of Mel Spectrograms.

### B. Training

All five experiments were done using ResNet18 as the model with cross entropy as the loss function and Adam as the optimizer. The datasets were transformed into tensors and then normalized to a mean of [0.485, 0.456, 0.406] and a standard deviation of [0.229, 0.224, 0.225], which are the same values used in the Imagenet1k_v1 weights. The dataset was split into 8,599 tensors for training, 1,075 tensors for validation during training, and 1074 tensors for final testing. The training happened over 20 epochs with a batch size of 32 tensors. The model was trained to classify the tensors as either deepfake or human.

### C. Outcomes

As would be expected, it was easiest for ResNet18 to detect deepfakes when they were unmodified. The baseline dataset gave the best results in most scores. The only time it was beaten was by the experiment I classified as "Continuous Learning" where I trained the same ResNet18 model on each dataset sequentially. The Continuous Learning model had the highest Recall score with 88.89%, beating all the other models at accurately detecting when a song was human. It also had the best False Negative Rate (FNR) of 11.11%. However, the continuous learning model did worse than all of the other models that were trained on a single dataset in every other measure, notably having double the second-worst False Positive Rate (FPR). The other models that were trained on one type of dataset were each within 3-4 percentage points of the Baseline experiment. An outcome of interest is that it appears easier to detect rudimentary pitch shifts compared to temporal stretches. Table I contains the metrics for each data source type.

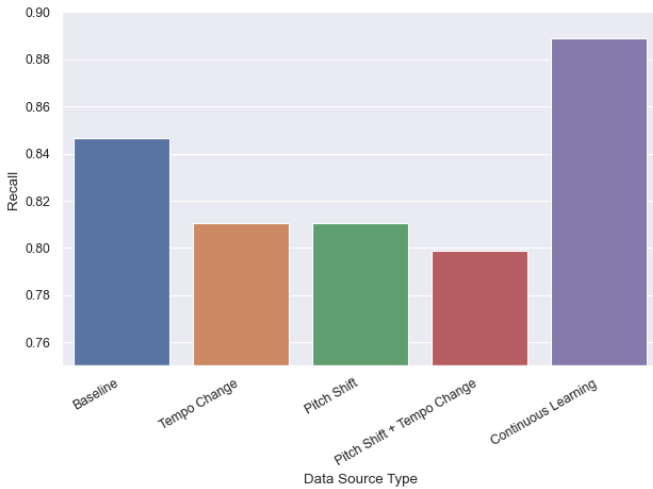

Fig. 1. Recall by Data Source Type

### D. Conclusions and Future Opportunities

The results indicate that the problem of adversarial techniques to evade deepfake detection is not an insurmountable one. I believe the Continuous Learning experiment had a lower FNR and higher Recall because it was able to reinforce what makes the audio "human" even with the modified datasets. With only 20 training epochs all models had scores over 80% for F1, Accuracy, Recall, and Precision which is quite remarkable.

Future research could take this same approach further in several areas. One area of research that would be of interest is how other models might perform with these same datasets, such as the deeper convolutional neural network ResNet50.

Another future area of experimentation that would be worth exploring is performing more extreme modifications to the audio datasets. The changes made to the datasets for this set of experiments was modelled after those done in Deezer's experiments [4] that a moderately skilled malicious actor could easily perform. Other forms of manipulation such as adding effects, e.g. reverb, echo, or distortion; or modifying the volume of the track would be good subjects to train the models on.

Two final opportunities for future experimentation would be to train the models over more epochs and use a larger sample of deepfake audio files. Expanding to the full set of 27,668 deepfake audio clips from the FakeMusicCaps dataset would create an even more robust split for training and testing. Training the model over more epochs would also allow it to draw more connections and gain greater performance.

## V. DISCUSSION OF MUSICAL DEEPFAKES

In this section I would like to discuss ethical and societal implications of the continued proliferation of TTM platforms. The ease of use and variety of platforms combined with the murky legal ground of the music generated by AI leads to some unique situations for the future of the music industry and society.

### A. Music Industry Impacts

We have already started to see various impacts on the music industry over the last few years. A song called *Heart on My Sleeve* went viral briefly before the record label of the imitated artists', Drake and the Weeknd, got involved to remove it from the internet [8]. It used a generative AI-empowered vocal encoder to create realistic mimicry of the two popular artists. Musician and Producer Rick Beato posits that music labels are going to capitalize on AI generated music from deceased musicians [10]. In a more recent video Beato suggests that AI generated music will cause the music industry as we know it to cease to exist [11]. Universal Media Group (UMG) has admitted they have been working on their "own innovation around AI for some time already [12]." Other popular musicians have been found to have deepfake models on popular AI music and vocal sites [13]. Some of these musicians are deceased, while others would likely prefer to keep their vocal likeness to themselves. Even those that consent to using their voice or likeness to train AI models might face legal issues over who owns the resulting work [14].

TABLE I
RESULTS BY DATA SOURCE

| Data Source Type | F1 Score | Accuracy | Recall | False Positive Rate | False Negative Rate | Precision | Specificity |
|---|---|---|---|---|---|---|---|
| Baseline | 0.878 | 0.885 | 0.847 | 0.078 | 0.153 | 0.911 | 0.922 |
| Tempo Stretch | 0.843 | 0.854 | 0.810 | 0.105 | 0.190 | 0.879 | 0.895 |
| Pitch Shift | 0.856 | 0.868 | 0.810 | 0.078 | 0.190 | 0.907 | 0.922 |
| Pitch Shift + Tempo Stretch | 0.837 | 0.844 | 0.799 | 0.105 | 0.201 | 0.878 | 0.895 |
| Continuous Learning | 0.842 | 0.835 | 0.889 | 0.217 | 0.111 | 0.800 | 0.783 |

### B. Legal Ramifications

The artist Drake has been at the forefront of legal issues surrounding AI recreations of artists, not only because he himself has been mimicked but because he has used AI to revive Tupac for a posthumous feature on a track. Drake did not ask for permission from the late rapper's estate, however, and was forced to remove the song from streaming platforms [15]. This is not the first time that the deceased rapper Tupac has been resurrected for entertainment purposes. His estate approved of a live performance using state of the art visual effects and holographic projection that happened at the Coachella festival in 2012 [16][17].

Listeners of Drake's song *Taylor Made Freestyle* in 2024, and concert goers back in 2012, both had reservations about reviving the rapper, who passed away from blood loss due to a drive-by shooting back in 1996. Some legal scholars even argue that posthumous deepfakes cause "dignitary and property harms [18]." These posthumous creations threaten artists reputations, which the deceased have a legal right to. Additionally, The Copyright Act of 1976 offers protections to musicians, composers, and publishers. These protections include the right to new works; the right to make, sell and distribute copies; and the right to publicly perform their works [19]. By training AI generation models on artists, living and deceased, those models are potentially infringing upon those rights.

The right of publicity is another legal right that could be infringed upon by deepfakes. It protects an individual's right to profit from their own personality as well as providing relief for unauthorized use of one's identity [20]. This could provide legal recourse for artists who find their vocal or physical likeness has been used to train AI models. Defamation, extortion, identity theft, reputational damage, and fraud are all other possible legal issues caused by the proliferation of deepfakes [21].

Intention to commit fraud is another illegal activity that has become possible due to the ease of creating musical deepfakes. A music producer was charged with multiple felonies related to intentionally scamming over $10 million worth of royalties from music streaming platforms [22]. In short, he intentionally devised a way to get around the platforms' fraud detection systems to upload massive amounts of songs. This is not the only occurrence of a scheme to rake in royalties. Earlier in 2024, thousands of fake albums were found on Spotify that were attributed to real artists [23]. The Ars Technica article also notes that while Spotify is a well-known streaming service with fraud detection built into their platform, typically these fake albums are discovered by music fans rather than their fraud detection systems. This highlights the need to continue research into developing robust deepfake detection models.

### C. Ethical Concerns

Julia Barnett found, in an extensive literature review of generative audio models, that less than 10% of researchers in the field discuss any negative impacts of their work [24]. Those that do mention negative impacts bring up several of the concerns mentioned in the previous section, *Legal Ramifications*. Some researchers noted lack of autonomy, agency, and authorship as an area for concern with AI generated material [25]. Plagiarism, musical identity dilution, decreasing need for human composers, and causing an erosion of culturally diverse human music-making traditions are concerns identified by Dimple Patil in their research [26].

A lack of diversity in datasets and the teams developing AI generative models has been noted by many different studies [27][28][29]. This results in AI models being biased to reflect the music and culture of the developers creating the models. While this is not a new issue in computer science, AI models compound the issue since they learn from historical data, essentially reinforcing these historical biases [30]. Biases in AI and machine learning (ML) can also arise from the way data is processed algorithmically. Measurement bias, omitted variable bias, representation bias, and aggregation bias are examples that can be present in ML training algorithms which might lead to biased outcomes [31].

The environmental impacts of AI generative audio models are not widely explored. Douwes, et al. discuss the need for a new method of evaluating the performance of models by also considering their energy consumption [32]. A different group of researchers found that the algorithm used to train AI has been shown to have an impact on the overall energy use of the AI platform [33]. This means that developers working with AI can have a positive, or less negative, environmental impact by choosing the correct algorithm for their work.

Falsely identifying human-made music as fake should be an area of concern for anyone trying to build a deepfake detector. The band Appalachian Anarchy became popular quickly in early 2024. They are primarily the passion project of a single musician, Peter Coffin. The band uses computer effects and MIDI instruments to create their music. Coffin also employs AI image generation software to create his album artwork [34]. The computer processing combined with a lack of live shows and AI artwork for the albums lead to the band being accused of being completely AI generated music [35]. The band claims their music has always been human-created and have even released videos of their own showing the creation process behind their music [36]. Automated fraud and deepfake detection systems will need to be vetted very seriously before they can be fully trusted to not harm the livelihoods and reputations of legitimate musicians.

### D. Positive Impacts

Deepfake generation technology has positive aspects as well. For example, there is a corporation named CereProc that uses deepfake technology to create artificial voices for people who have lost theirs from disease [37]. Similarly, singer Randy Travis suffered a major stroke in 2013 and lost the ability to use his voice. Warner Music, Travis, and his long-time music producer worked with an AI company to artificially recreate his voice to create a new song [38]. The ability to allow someone to speak or sing with their own voice that was lost due to health-related issues is a great use of deepfake technology that can help restore some semblance of bodily autonomy. The amount of people that can be helped with these sorts of projects will only grow as the technology matures and becomes more readily available.

The Beatles made waves in 2023 by employing AI to finish and release the track *Now and Then*. It posthumously featured John Lennon whose vocals, originally recorded in his home in 1977, were restored using AI. Human artists are also using AI music generation platforms as a tool to brainstorm ideas. They can put the style of music they're interested in making in the TTM prompt and riff off the music the model creates. Some artists, like Claire L. Evans of the band YACHT, even say that when the TTM platform creates a "failure" it can be strange and work to get the creative juices flowing even more than a "successful" creation [39]. Creative humans are using AI streaming platforms similarly to how artists would sample songs in the past. In the summer of 2024, artist Metro Boomin went viral for his instrumental track *BBL Drizzy* which he created using samples from a song created by comedian Willonius Hatcher on the AI music platform Udio [40].

Another way that humans have been using AI TTM platforms to help them is by creating instrumental stock music tracks. These can be placeholders while human composers create the final projects for films, or they could even be used by students for class projects. Creative humans are also using AI generated music tools to create a sort of "alternate reality" mashup. One such artist that has gained viral fame is Dustin Ballard, who goes by the moniker There I Ruined It. One popular example of his style of mashup is Jay-Z's 99 Problems "performed" by the Beach Boys [41]. AI music models can also be helpful to mashup artists by splitting songs into their music stems. These can then be spliced together from various sources to create a new mashup song.

### E. Final Thoughts

Musical deepfakes and the platforms to generate them are not going away any time soon. They pose a lot of legal and ethical questions we need to address as a society, but they also create a lot of opportunities for creativity and creative problem solving. AI powered deepfake detectors will play a part in keeping artists safe from impersonation and fraud if they are employed in a thoughtful and ethical way.

### ACKNOWLEDGMENT

I appreciate the FakeMusicCaps team which laid a great foundation for future work with deepfake detection. I am also grateful for the maintainers and documenters of the PyTorch [42] and Librosa [43] Python libraries which were crucial for this project.